\documentclass[a4paper,twocolumn,american]{revtex4}
\usepackage{mathptmx}
\usepackage[T1]{fontenc}
\usepackage[latin1]{inputenc}
\usepackage{amsmath}
\usepackage{graphicx}

\makeatletter

\usepackage{amsmath}
\usepackage{graphicx}
\usepackage{amssymb}

\usepackage{babel}

\makeatletter

\let\citeleft=(
\let\citeright=)

\renewcommand\@biblabel[1]{#1.} 

\def\@cite#1#2{\leavevmode \cite@adjust
  \citeleft{\it {#1\if@tempswa \citemid #2\fi
  \spacefactor\@m 
  }}\citeright}

\makeatother

\makeatother

\usepackage{babel}

\begin{document}

\title{Towards a loophole-free test of Bell's inequality with entangled
pairs of neutral atoms}

\author{Wenjamin Rosenfeld$^{1}$, Markus Weber$^{1}$, Jürgen Volz$^{2}$,
Florian Henkel$^{1}$, Michael Krug$^{1}$, \\
Adán Cabello$^{3}$, Marek \.{Z}ukowski$^{4}$, and Harald Weinfurter$^{1,5}$}

\affiliation{$^{1}$Fakultät für Physik, Ludwig-Maximilians Universität München,
D-80799 München, Germany}

\email{w.r@lmu.de}

\affiliation{$^{2}$Laboratoire Kastler Brossel de l'E.N.S., F-75005 Paris, France}

\affiliation{$^{3}$Departamento de Fisica Aplicada II, Universidad de Sevilla,
E-41012 Sevilla, Spain}

\affiliation{$^{4}$Instytut Fizyki Teoretycznej i Astrofizyki, Uniwersytet Gdanski,
PL-80-952 Gdansk, Poland}

\affiliation{$^{5}$Max-Planck Institut für Quantenoptik, D-85748 Garching, Germany}
\begin{abstract}
Experimental tests of Bell's inequality allow to distinguish quantum
mechanics from local hidden variable theories. Such tests are performed
by measuring correlations of two entangled particles (e.g. polarization
of photons or spins of atoms). In order to constitute conclusive evidence,
two conditions have to be satisfied. First, strict separation of the
measurement events in the sense of special relativity is required
({}``locality loophole''). Second, almost all entangled pairs have
to be detected (for particles in a maximally entangled state the required
detector efficiency is $2(\sqrt{2}-1)\approx82.8\%$), which is hard
to achieve experimentally ({}``detection loophole''). By using the
recently demonstrated entanglement between single trapped atoms and
single photons it becomes possible to entangle two atoms at a large
distance via entanglement swapping. Combining the high detection efficiency
achieved with atoms with the space-like separation of the atomic state
detection events, both loopholes can be closed within the same experiment.
In this paper we present estimations based on current experimental
achievements which show that such an experiment is feasible in future.
\end{abstract}
\maketitle

\section{Introduction}

In 1935 Einstein, Podolsky and Rosen (EPR) asked the seemingly innocent
question, whether quantum mechanics can be considered complete. If
not, this might be cured by additional parameters of a physical system
(now called local hidden variables, LHV) which are not - yet - known
to us. Later, Bell showed, that experimental tests can be performed
which allow to decide whether the concept of LHV indeed can be used
to describe nature. This proposal triggered a series of experiments,
most importantly by Freedman \& Clauser\cite{Freedman72} and by the
group of Alain Aspect\cite{Aspect82a,Aspect82b}. More recently, new
experimental techniques enabled Bell-tests with photon pairs from
parametric down-conversion and, with the realm of quantum logic, for
trapped ions, nuclear spins etc.

So far, all experiments to test Bell's inequalities required additional
assumptions, thus opening loopholes in Bell's original argument\cite{Clauser78}.
The first is called the locality loophole, in which the correlations
of apparently separate events could result from unknown subluminal
signals influencing the measurement results during the observation
of an entangled pair\cite{Bell88,Santos92}. One experiment was performed
with entangled photons\cite{Weihs98} enforcing strict relativistic
separation between the measurements. But it suffered from low detection
efficiencies. It thus opens the second loophole by allowing the possibility
that the subensemble of detected events agrees with quantum mechanics
even though the entire ensemble satisfies the limits for local-realistic
theories as given by Bell's inequalities\cite{Pearle70,Santos95}.
This is also referred to as detection loophole and was addressed in
an experiment with two trapped ions\cite{Matsukevich08}, where the
quantum state detection was performed with almost perfect efficiency.
But there the ion separation was too small to eliminate the locality
loophole.

\begin{figure}
\begin{centering}
\includegraphics{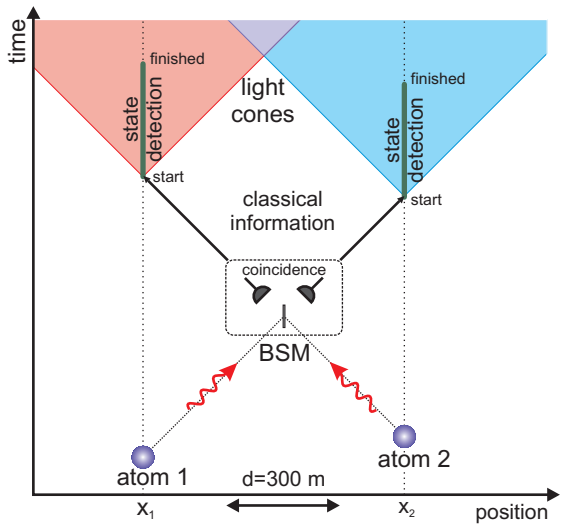}
\par\end{centering}

\caption{\label{fig:SpaceTimeDiagram}Space-time schematic of the proposed
loophole-free Bell experiment. Two atomic traps are separated by $300\,\mathrm{m}$,
each atom emits a photon whose polarization is entangled with the
atomic spin. The two photons arrive simultaneously on a non-polarizing
beamsplitter where interference takes place. The coincidence detection
in the outputs of the beamsplitter (equivalent to a Bell-state measurement
(BSM) on the two photons) signals the projection of the atoms onto
an entangled state. The signal of successful BSM is sent back to both
setups, where atomic state detection is started. The detection is
performed in a randomly chosen basis and has to be finished before
any classical signal can reach the other side (i.e. within less than
$1\,\mathrm{\mu s}$).}

\end{figure}

Based on the experiments performed in our group\cite{Volz06,longDistAPE},
a final test of LHV-theories\cite{Z} comes into reach of our experimental
techniques. For this purpose two photons, each entangled with a trapped
$^{87}Rb$ atom, will be distributed far enough to ensure space-like
separation, see Fig. \ref{fig:SpaceTimeDiagram}. A projection of
the photons onto Bell-states serves to swap the entanglement to the
atoms\cite{Zukowski93} whose states now can be observed with high
efficiency. This enables the ideal configuration of a so called event-ready
scheme\cite{Clauser78,Bell88,Zukowski93}, which does not require
any assumptions at all.

\section{Experimental requirements}

Let us now analyze the experimental requirements. Crucial for such
a test is a highly efficient state analysis performed by space-like
separated observations on entangled atoms. Here the minimum distance
between the atoms is determined by the duration of the atomic state
detection process.

The currently used atomic state detection method is a two-step process\cite{Volz06}.
It consists of a stimulated Raman adiabatic passage technique (STIRAP)
which transfers a selected superposition of the atomic spin states
to a different hyperfine level ($F=2$) and a subsequent detection
of the hyperfine state. While the STIRAP process is inherently coherent,
the coherence of the atomic state is destroyed right after the STIRAP
sequence by resonant scattering of photons within $300\,\mathrm{ns}$
with a probability exceeding $99\%$. Alternatively, the hyperfine
state detection can be replaced by state-selective ionization with
subsequent detection of the ionization fragments. By irreversibly
removing the valence electron, the coherence of the atom is destroyed
(according to calculations) after $200\,\mathrm{ns}$ with a probability
of $>99\%$. Together with the random choice of the measurement basis
($100\,\mathrm{ns}$), the STIRAP process ($120\,\mathrm{ns}$), and
flight times of the ionization fragments ($<500\,\mathrm{ns}$) it
gives an overall detection time of less than $1\,\mathrm{\mu s}$.
The corresponding distance of $300\,\mathrm{m}$ between the atoms
for closing the locality loophole can easily be achieved since the
transmission losses in optical fibers for the photons used for entanglement
swapping (wavelength $780\,\mathrm{nm}$) are low (for a demonstration
of an optical fiber link of $300\,\mathrm{m}$ length see\cite{longDistAPE}).
We emphasize that our scheme is also independent of any detection
related loopholes, because entanglement swapping enables the event
ready scheme\cite{Clauser78,Bell88,Zukowski93}, where binary measurement
results are reported for every run, started after a joint photon detection
event in the Bell-state measurement. For limited detection efficiency/accuracy,
however, the obtained results are not always correct. This leads to
a reduction of the expected spin correlations. The corresponding accuracies
of the two detection methods are analyzed in this paper and the expected
violation of Bell's inequality is given.

\subsection{State-selective atom removal}

The currently used detection of the hyperfine state involves state-selective
removal of the atoms from the trap, which is verified by counting
photons collected from the trap region. The mean accuracy of this
procedure was experimentally determined to be $a_{HF}=97.8\%$\cite{VolzThesis}.
Together with the accuracy of the STIRAP process, $a_{ST}=97.25\%$
it results in an overall detection accuracy of $a_{det}^{(flr.)}=95\%$.
This number specifies the (symmetric) probability for correct identification
of the analyzed atomic state (i.e. $\left|\downarrow\right\rangle $
is identified as $\left|\downarrow\right\rangle $ and $\left|\uparrow\right\rangle $
as $\left|\uparrow\right\rangle $). A disadvantage of this method
due to very low collection efficiency of only about $10^{-3}$ is
the long duration of sampling fluorescence photons until the outcome
can be determined ($10..20\,\mathrm{ms}$). Yet, one should note that
decoherence (coupling to the environment) already takes place within
short time ($300\,\mathrm{ns}$) by scattering a single photon.

\subsection{State-selective ionization }

Alternatively, in order to enable a very fast and direct detection
of the atomic state, state-selective ionization can be used. Here
again a selected superposition of atomic spin states is first transferred
to $5^{2}S_{3/2},\, F=2$ hyperfine level using the STIRAP technique.
Then the atom in $F=2$ level is optically excited to the $5^{2}P_{3/2},\: F=3$
level and ionized using an additional laser at a wavelength of $473\,\mathrm{nm}$.
The rate of this two-photon ionization process depends on the available
intensity of the lasers. We expect to achieve an ionization probability
of $p_{ionize}>99\%$ within $200\,\mathrm{ns}$. The resulting free
electron $e^{-}$ and Rubidium ion $^{87}Rb^{+}$ can be detected
by channel electron multipliers. As it is sufficient to detect at
least one of the ionization fragments, the overall detection efficiency
$p_{det}$ is given by\begin{equation}
p_{det}=1-(1-p_{e})(1-p_{ion}).\label{eq:ionizationDetEff}\end{equation}
 This method is currently investigated in our group. First calibration
measurements for ionization of Rubidium atoms from background gas
in a vacuum cell show efficiencies of $p_{e}=80\%$ and $p_{ion}=60\%$.
The goal is to reach values $p_{e}\geq85\%$ and $p_{ion}\geq65\%$,
which would give a detection efficiency of $p_{det}=95\%$ and better.

Again it has to be stressed that the efficiency for detection of ionization
fragments is not the detection efficiency in the Bell experiment.
Due to the binary nature of the result (either a fragment is detected
corresponding to the measurement result {}``$\left|\uparrow\right\rangle $'',
or it is not detected, corresponding to the measurement result {}``$\left|\downarrow\right\rangle $'',
but a result is always given) this efficiency does only influence
the accuracy of the state detection.

\section{Expected visibility for the entanglement swapping}

For all further considerations we assume that the entangled state
of atom-photon or two atoms has the density matrix of the following
form\begin{equation}
\hat{\rho}=V\left|\Psi\right\rangle \left\langle \Psi\right|+(1-V)\frac{1}{4}\hat{1},\label{eq:densityMatrix}\end{equation}
 where $V$ is the visibility, $\left|\Psi\right\rangle =\frac{1}{\sqrt{2}}(\left|\downarrow\right\rangle \left|\uparrow\right\rangle \pm\left|\uparrow\right\rangle \left|\downarrow\right\rangle )$
is a maximally entangled state and $\frac{1}{4}\hat{1}$ is the density
matrix of the completely mixed state\cite{A}. In a correlation measurement,
where the relative angle between the measurement bases of the two
particles is varied, the visibility $V$ describes the difference
between the maximum and the minimum values (also called contrast)
of the observed interference fringe. Given the state represented by
the density matrix $\hat{\rho}$ from (\ref{eq:densityMatrix}), the
probability to find the two particles in the (pure) state $\left|\Psi\right\rangle $
(also called the fidelity $F$) is $F=\frac{1}{4}+\frac{3}{4}V$.

For any additional error occurring at the further stages of the experiment
we assume that the density matrix is modified like\[
\hat{\rho}\rightarrow(1-e)\hat{\rho}+e\cdot\frac{1}{4}\hat{1},\]
 where $e$ is the error probability. This assumes that any error
results in a completely mixed state. For visibility $V$ and fidelity
$F$ of the state follows\begin{alignat}{1}
V\rightarrow & (1-e)V,\nonumber \\
F\rightarrow & (1-e)F+\frac{1}{4}e.\label{eq:reductionVisFid}\end{alignat}
 These relations allow to calculate the influence of different errors
during the transmission of the state, entanglement swapping, etc.

In order to generate an entangled pair of atoms, the starting situation
is the emission of a photon by the atom. During this process the polarization
of the photon gets entangled with the respective atomic spin resulting
in the maximally entangled state\cite{Volz06}

\[
\left|\Psi^{+}\right\rangle _{at-ph}=\frac{1}{\sqrt{2}}(\left|\downarrow\right\rangle _{z}\left|\sigma^{+}\right\rangle +\left|\uparrow\right\rangle _{z}\left|\sigma^{-}\right\rangle ).\]
 The two states $\left|\uparrow\right\rangle _{z}$ and $\left|\downarrow\right\rangle _{z}$,
defining the atomic qubit, correspond to the $\left|F=1,\, m_{F}=\pm1\right\rangle $
Zeeman sublevels of the $5{}^{2}S_{1/2},\: F=1$ hyperfine ground
level. The purity of this state is limited only by the errors in preparation
of the excited state\cite{footnote1}, in our case we assume $e_{exc}=0.5\%$
due to imperfections in the preparation of the initial state and resulting
off-resonant excitation to different atomic states, leading to $V_{at-ph}^{(initial)}=99.5\%$.
The smaller visibility observed in the current experiments\cite{Volz06}
is due to errors in the analysis of the atom-photon state which are
described below. For the generation of atom-atom entanglement via
entanglement swapping, the photon propagates via an optical fiber
to a different location where the two-photon interference takes place.
Recently we have demonstrated an optical fiber link of $300\,\mathrm{m}$
length\cite{longDistAPE}, where the polarization errors were kept
below $1\%$ by active polarization control. Thus the remaining polarization
errors in the fiber ($e_{pol}=1\%$) reduce the visibility to \[
V_{at-ph}=(1-e_{exc})(1-e_{pol})=98.5\%.\]
 This is the atom-photon visibility which is assumed before the photons
enter the apparatus for the Bell-state measurement (BSM).

In the entanglement swapping process an additional error might occur
due to mismatch in the two-photon interference which is assumed to
be $e_{BSM}=3\%$. The projection of the two atoms onto the entangled
state is heralded by the coincidence detection (double click) of the
two photons leaving two different output ports of the beamsplitter.
Conditioned on this coincidence, the probability $p(\left|\Psi^{-}\right\rangle _{at-at})$
to get the desired entangled atom-atom state $\left|\Psi^{-}\right\rangle _{at-at}$
is\begin{equation}
(1-e_{BSM})(\frac{1}{4}+\frac{3}{4}V_{at-ph}^{2})+\frac{1}{4}e_{BSM}=95.6\%,\label{eq:fidelity-EntSwap}\end{equation}
 where the influence of the error $e_{BSM}$ follows from (\ref{eq:reductionVisFid}).

Dark counts in the single photon detectors of the Bell-state analyzer
will add spurious events. The fraction of wrong coincidence events
is calculated as follows. The probability to get a photon from the
first trapped atom at the beamsplitter is $\eta_{1}=1.3\cdot10^{-3}\times0.6=0.78\cdot10^{-3}$,
where the first number is the local efficiency for the generation
of entangled atom-photon pairs (including the detection efficiency
of single-photon detectors), while the second number accounts for
the coupling and transmission losses in the fiber, as well as the
limited time window for the coincidence detection. For the photon
from the second atomic trap this number is higher due to the higher
numerical aperture, $\eta_{2}=2.0\cdot10^{-3}\times0.6=1.2\cdot10^{-3}$.
Therefore the probability to detect a coincidence of the two photons
is\begin{equation}
p_{coincidence}^{(true)}=\frac{1}{4}\eta_{1}\eta_{2}=2.34\cdot10^{-7}.\label{eq:efficiency-EntSwap}\end{equation}
 The factor $\frac{1}{4}$ accounts for the fact that only one out
of four photonic Bell-states is detected. A {}``wrong'' coincidence
happens if one photon arrives at the beamsplitter and is detected
in one detector while the other detector produces a dark count within
the coincidence time window. For the detectors which will be used
for this purpose (Perkin-Elmer SPCM-AQR15) the dark count rate is
$r_{dc}\leq50\,\mathrm{cps}$. For a coincidence time window of $\Delta T=40\,\mathrm{ns}$
the probability of such an event is\[
p_{coincidence}^{(dark)}\approx(\eta_{1}+\eta_{2})r_{dc}\Delta T=3.96\cdot10^{-9}.\]
 As the probability of detecting two dark counts as coincidence is
negligible ($4\cdot10^{-12}$), the fraction of wrong events in the
coincidence detection is $e_{dc}=1.68\%$. Applying the relations
(\ref{eq:reductionVisFid}) to the fidelity from (\ref{eq:fidelity-EntSwap})
we obtain a resulting fidelity of $F_{at-at}=94.4\%$ and visibility
of $V_{at-at}=\frac{1}{3}(4F_{at-at}-1)=92.5\%$.

\section{Expected violation of Bell's inequality}

For the experimental test of the CHSH formulation of Bell's inequality,
the parameter $S$ is measured, which is defined as

\begin{equation}
S:=\left|\left\langle \sigma_{\alpha}\sigma_{\beta}\right\rangle +\left\langle \sigma_{\alpha'}\sigma_{\beta}\right\rangle \right|+\left|\left\langle \sigma_{\alpha}\sigma_{\beta'}\right\rangle -\left\langle \sigma_{\alpha'}\sigma_{\beta'}\right\rangle \right|.\label{eq:S-Parameter}\end{equation}
 Here $\left\langle \sigma_{\alpha}\sigma_{\beta}\right\rangle $
is the expectation value of joint measurements on the spins of two
particles where one spin is analyzed at an angle $\alpha$ and the
other one at an angle $\beta$ (we define these angles in terms of
light polarization in the laboratory frame). According to Bell's theorem,
any theory with local hidden variables predicts $S\leq2$. In quantum
mechanics $S=2\sqrt{2}$ is reached, e.g. for $\alpha=0^{\circ}$,
$\alpha'=45^{\circ}$, $\beta=22.5^{\circ}$, $\beta'=-22.5^{\circ}$.

In an experiment we measure the number of events {}``$\uparrow\uparrow$'',
{}``$\downarrow\downarrow$'', {}``$\uparrow\downarrow$'', {}``$\downarrow\uparrow$'',
where the {}``ups'' and {}``downs'' are the orientations of the
spins with respect to the corresponding analysis directions $\alpha,\beta$.
We shall call these numbers $N_{\uparrow\uparrow}^{(\alpha,\beta)}$,
$N_{\downarrow\downarrow}^{(\alpha,\beta)}$, $N_{\uparrow\downarrow}^{(\alpha,\beta)}$,
$N_{\downarrow\uparrow}^{(\alpha,\beta)}$, while the total number
of events per setting $(\alpha,\beta)$ is $N_{s}=N_{\uparrow\uparrow}^{(\alpha,\beta)}+N_{\downarrow\downarrow}^{(\alpha,\beta)}+N_{\uparrow\downarrow}^{(\alpha,\beta)}+N_{\downarrow\uparrow}^{(\alpha,\beta)}$.
The expectation values are calculated as \begin{multline}
\left\langle \sigma_{\alpha}\sigma_{\beta}\right\rangle =\frac{1}{N_{s}}(N_{\uparrow\uparrow}^{(\alpha,\beta)}+N_{\downarrow\downarrow}^{(\alpha,\beta)}-N_{\uparrow\downarrow}^{(\alpha,\beta)}-N_{\downarrow\uparrow}^{(\alpha,\beta)})\\
=\frac{2}{N_{s}}(N_{\uparrow\uparrow}^{(\alpha,\beta)}+N_{\downarrow\downarrow}^{(\alpha,\beta)})-1.\label{eq:correlator}\end{multline}
 We note that\begin{alignat}{1}
N_{\uparrow\uparrow}^{(\alpha,\beta)}= & N_{s}\cdot p_{\uparrow\uparrow}^{(\alpha,\beta)},\nonumber \\
N_{\downarrow\downarrow}^{(\alpha,\beta)}= & N_{s}\cdot p_{\downarrow\downarrow}^{(\alpha,\beta)},\label{eq:eventN}\end{alignat}
 where\begin{alignat*}{1}
p_{\uparrow\uparrow}^{(\alpha,\beta)}= & p\left(\left|\uparrow\right\rangle _{1}^{(\alpha)}\left|\uparrow\right\rangle _{2}^{(\beta)}\right)\\
p_{\downarrow\downarrow}^{(\alpha,\beta)}= & p\left(\left|\downarrow\right\rangle _{1}^{(\alpha)}\left|\downarrow\right\rangle _{2}^{(\beta)}\right)\end{alignat*}
 are the probabilities for both particles to be measured in the state
$\left|\uparrow\right\rangle $ ($\left|\downarrow\right\rangle $)
along their respective analysis direction. For the atomic states the
relations \begin{alignat*}{1}
\left|\uparrow\right\rangle ^{(\alpha)}= & \cos(\beta-\alpha)\left|\uparrow\right\rangle ^{(\beta)}+\sin(\beta-\alpha)\left|\downarrow\right\rangle ^{(\beta)}\\
\left|\downarrow\right\rangle ^{(\alpha)}= & \cos(\beta-\alpha)\left|\downarrow\right\rangle ^{(\beta)}-\sin(\beta-\alpha)\left|\uparrow\right\rangle ^{(\beta)}\end{alignat*}
 hold and therefore\begin{multline*}
\left|\Psi^{-}\right\rangle =\frac{1}{\sqrt{2}}(\left|\downarrow\right\rangle _{1}^{(\alpha)}\left|\uparrow\right\rangle _{2}^{(\alpha)}-\left|\uparrow\right\rangle _{1}^{(\alpha)}\left|\downarrow\right\rangle _{2}^{(\alpha)})\\
=\frac{1}{\sqrt{2}}\Big(\cos(\beta-\alpha)\left|\uparrow\right\rangle _{1}^{(\alpha)}\left|\downarrow\right\rangle _{2}^{(\beta)}-\sin(\beta-\alpha)\left|\uparrow\right\rangle _{1}^{(\alpha)}\left|\uparrow\right\rangle _{2}^{(\beta)}\\
-\cos(\beta-\alpha)\left|\downarrow\right\rangle _{1}^{(\alpha)}\left|\uparrow\right\rangle _{2}^{(\beta)}-\sin(\beta-\alpha)\left|\downarrow\right\rangle _{1}^{(\alpha)}\left|\downarrow\right\rangle _{2}^{(\beta)}\Big).\end{multline*}
 The probabilities $p_{\uparrow\uparrow}^{(\alpha,\beta)}$, $p_{\downarrow\downarrow}^{(\alpha,\beta)}$
are explicitely calculated in the following by applying the experimental
detection probabilities and accuracies depending on the detection
method.

\subsection{Atomic state analysis via state-selective atom removal and fluorescence
detection}

When the entangled atom-atom state (\ref{eq:densityMatrix}) with
an initial visibility $V_{at-at}$ is analyzed, we expect the probabilities\begin{multline}
p_{\uparrow\uparrow}^{(\alpha,\beta)}=p_{\downarrow\downarrow}^{(\alpha,\beta)}\\
=\frac{1}{4}\left(1-V_{at-at}(2a_{det}-1)^{2}\cos(2(\beta-\alpha))\right).\label{eq:prob-FlrDet}\end{multline}
 of detecting both particles in the state $\left|\uparrow\right\rangle $,
respectively $\left|\downarrow\right\rangle $ along the directions
$(\alpha,\beta)$. Inserting this into (\ref{eq:S-Parameter}, \ref{eq:correlator},
\ref{eq:eventN}) we determine the expected parameter $S$\begin{equation}
S^{(flr.)}=2\sqrt{2}V_{at-at}(2a_{det}-1)^{2}.\label{eq:S-FlrDet}\end{equation}
 For $V_{at-at}=92.5\%$, $a_{det}^{(flr.)}=95\%$ this gives $S^{(flr.)}=2.12$,
corresponding to an observable atom-atom visibility of $V^{(flr.)}=74.9\%$.

\subsection{Atomic state analysis via state-selective ionization}

The limited detection efficiency for the ionization fragments leads
to an asymmetry in the accuracy for the two measurement outcomes.
The result where one of the channel electron multipliers registers
a particle definitely means that an ionization has taken place (the
probability of a dark count is low and therefore neglected). However,
the result where no particle is registered contains also the events
where the ionized fragments were not detected\cite{Garg87}. The probabilities
in this case are\begin{alignat}{1}
p_{\uparrow\uparrow}^{(\alpha,\beta)}= & \frac{1}{4}p_{d}^{2}\left(1-V_{at-at}(2a_{ST}-1)^{2}\cos(2(\beta-\alpha))\right)\nonumber \\
p_{\downarrow\downarrow}^{(\alpha,\beta)}= & \frac{1}{4}(2-p_{d})^{2}\times\nonumber \\
 & \times\left(1-\frac{p_{d}^{2}}{(2-p_{d})^{2}}V_{at-at}(2a_{ST}-1)^{2}\cos(2(\beta-\alpha))\right),\label{eq:prob-IonizDet}\end{alignat}
 where we have set $p_{d}=p_{ionize}\cdot p_{det}$ for brevity. The
parameter $S$ is then given by\begin{equation}
S^{(ioniz)}=2\sqrt{2}V_{at-at}p_{d}^{2}(2a_{ST}-1)^{2}-2(1-p_{d})^{2}.\label{eq:S-ionizDet}\end{equation}
 This expression is exactly valid for $p_{d}\geq(1+\sqrt[4]{2})^{-1}$.
For the parameters $V_{at-at}=92.5\%$, $a_{ST}=97.25\%$, $p_{d}=95\%$
we get $S^{(ioniz)}=2.10$.

\section{Statistical uncertainty for the violation of Bell's inequality}

In order to violate Bell's inequality the value of $S>2$ has to be
measured with sufficient statistical significance. Calling the standard
deviation of the measured value $\Delta S$, it has to be assured
that \begin{equation}
\frac{S-2}{\Delta S}\geq k,\label{eq:BellViolation}\end{equation}
 where $k$ is the number of standard deviations for the violation.
Taking $k=3$ gives a confidence level of $\geq99.73\%$. The standard
deviation $\Delta S$ depends on the number of measured events and
shall be calculated in the following.

Using Gaussian error propagation we get from (\ref{eq:correlator})\[
\Delta\left\langle \sigma_{\alpha}\sigma_{\beta}\right\rangle =\frac{2}{N_{s}}\sqrt{\Delta N_{\uparrow\uparrow}^{2}+\Delta N_{\downarrow\downarrow}^{2}}.\]
 The uncertainty of $S$ is\begin{equation}
\Delta S=\sqrt{\sum_{\alpha,\beta}\Delta\left\langle \sigma_{\alpha}\sigma_{\beta}\right\rangle ^{2}},\label{eq:error-S}\end{equation}
 where $\alpha=0^{\circ},45^{\circ},\,\beta=22.5^{\circ},-22.5^{\circ}$.

Next, the statistical uncertainties of the event numbers have to be
determined. Here we note that for a Bernoulli experiment the standard
deviation of the expectation value is given by \begin{alignat}{1}
\Delta N_{\uparrow\uparrow} & =\sqrt{N_{\uparrow\uparrow}p_{\uparrow\uparrow}(1-p_{\uparrow\uparrow})}=\sqrt{N_{s}}\sqrt{p_{\uparrow\uparrow}^{2}(1-p_{\uparrow\uparrow})},\nonumber \\
\Delta N_{\downarrow\downarrow} & =\sqrt{N_{\downarrow\downarrow}p_{\downarrow\downarrow}(1-p_{\downarrow\downarrow})}=\sqrt{N_{s}}\sqrt{p_{\downarrow\downarrow}^{2}(1-p_{\downarrow\downarrow})}.\label{eq:error-EventN}\end{alignat}
 With these expressions the uncertainty of the $S$ parameter is calculated
for the two considered detection methods.

\subsection{Fluorescence detection}

Using the expression (\ref{eq:prob-FlrDet}) and taking the specific
angles for the Bell measurement we obtain

\[
p_{\uparrow\uparrow}^{(\alpha,\beta)}=p_{\downarrow\downarrow}^{(\alpha,\beta)}=\frac{1}{4}(1\mp\frac{1}{\sqrt{2}}V),\]
 where $V=V_{at-at}(2a_{det}-1)^{2}$, the {}``-'' sign is valid
for the settings $(0^{\circ},\pm22.5^{\circ})$, $(45^{\circ},22.5^{\circ})$
while the {}``+'' sign appears in the setting $(45^{\circ},-22.5^{\circ})$.
This expression is inserted into (\ref{eq:error-EventN}) giving\begin{multline*}
\Delta N_{\uparrow\uparrow}=\Delta N_{\downarrow\downarrow}\\
=\frac{\sqrt{N_{s}}}{4}\sqrt{(1\mp\frac{1}{\sqrt{2}}V)^{2}(1-\frac{1}{4}(1\mp\frac{1}{\sqrt{2}}V))},\end{multline*}
 Therefore for $(\alpha,\beta)$ equal to $(0^{\circ},\pm22.5^{\circ})$
and $(45^{\circ},22.5^{\circ})$\[
\Delta\left\langle \sigma_{\alpha}\sigma_{\beta}\right\rangle =\frac{1}{2\sqrt{2}\sqrt{N_{s}}}\sqrt{(1-\frac{1}{\sqrt{2}}V)^{2}(3+\frac{1}{\sqrt{2}}V)}\]
 and for $(\alpha,\beta)$ equal to $(45^{\circ},-22.5^{\circ})$\[
\Delta\left\langle \sigma_{\alpha}\sigma_{\beta}\right\rangle =\frac{1}{2\sqrt{2}\sqrt{N_{s}}}\sqrt{(1+\frac{1}{\sqrt{2}}V)^{2}(3-\frac{1}{\sqrt{2}}V)}.\]
 Using (\ref{eq:error-S}) we finally get

\begin{multline}
\Delta S^{(flr.)}=\frac{1}{\sqrt{2}\sqrt{N}}\times\\
\times\sqrt{3(1-\frac{1}{\sqrt{2}}V)^{2}(3+\frac{1}{\sqrt{2}}V)+(1+\frac{1}{\sqrt{2}}V)^{2}(3-\frac{1}{\sqrt{2}}V)},\label{eq:error-S-Flr}\end{multline}
 where $N=4N_{s}$ is the total number of events for all four settings
together.

\begin{figure}
\begin{centering}
\includegraphics{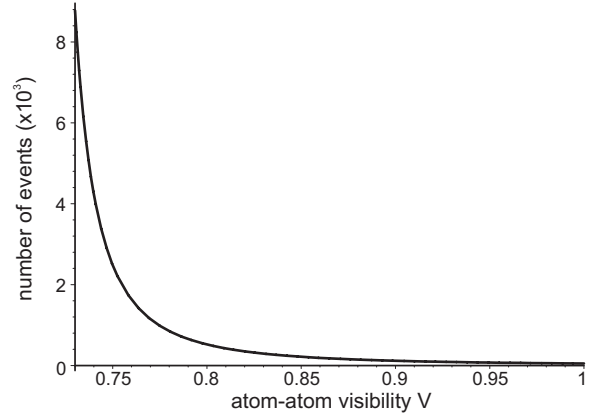}
\par\end{centering}

\caption{\label{cap:NvsV-FluorDet}Number $N$ of events necessary to violate
Bell's inequality by 3 standard deviations using fluorescence detection
as a function of the expected atom-atom visibility $V=V_{at-at}(2a_{det}-1)^{2}$.}

\end{figure}

Inserting this result into the expression for violation of Bell's
inequality (\ref{eq:BellViolation}) we can estimate the number of
events necessary to achieve a certain confidence level. Figure \ref{cap:NvsV-FluorDet}
shows the dependence of the number of events $N$ for a violation
by $3$ standard deviations as a function of the expected atom-atom
visibility $V=V_{at-at}(2a_{det}-1)^{2}$. For a visibility of $V=74.9\%$
we get $N=2600$.

\subsection{Ionization detection}

\begin{figure}
\begin{centering}
\includegraphics{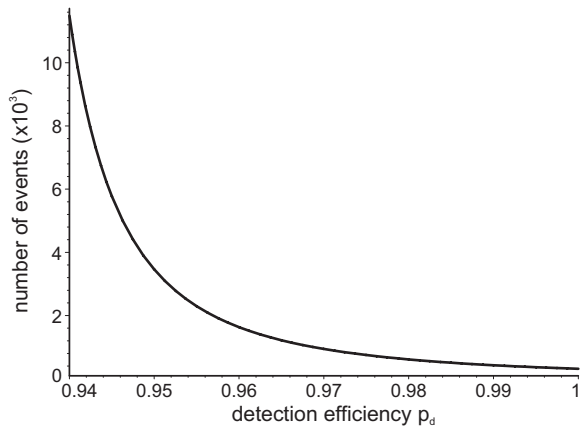}
\par\end{centering}

\caption{\label{cap:NvsPd-IonizDet}Number $N$ of events necessary to violate
Bell's inequality by 3 standard deviations with ionization detection
as a function of the electron/ion detection efficiency $p_{d}$ (including
ionization probability). The assumed atom-atom visibility excluding
the ionization detection efficiency is $V_{at-at}(2a_{ST}-1)^{2}=82.6\%$.}

\end{figure}

Using the expression (\ref{eq:prob-IonizDet}) and taking the specific
angles for the Bell measurement we obtain

\begin{alignat*}{1}
p_{\uparrow\uparrow}^{(\alpha,\beta)}= & \frac{1}{4}p_{d}^{2}\left(1\mp\frac{1}{\sqrt{2}}V_{at-at}(2a_{ST}-1)^{2}\right),\\
p_{\downarrow\downarrow}^{(\alpha,\beta)}= & \frac{1}{4}(2-p_{d})^{2}\left(1\mp\frac{1}{\sqrt{2}}\frac{p_{d}^{2}}{(2-p_{d})^{2}}V_{at-at}(2a_{ST}-1)^{2}\right),\end{alignat*}
 where $p_{d}=p_{ionize}\cdot p_{det}$. Again the {}``-'' sign
is for the settings $(0^{\circ},\pm22.5^{\circ})$, $(45^{\circ},22.5^{\circ})$
while the {}``+'' sign appears in the setting $(45^{\circ},-22.5^{\circ})$.
These are used for calculation of the uncertainty of the $S$ parameter
similar to the previous section. It is again inserted into (\ref{eq:BellViolation})
to estimate the necessary number of events. Figure \ref{cap:NvsPd-IonizDet}
shows the dependence of the required number of events on the detection
efficiency. Here we have assumed $V_{at-at}(2a_{ST}-1)^{2}=82.6\%$.
For the detection efficiency $p_{d}=95\%$ we get $N=3470$ events.

\section{Experimental event rates and measurement time}

In this section we estimate the repetition rate of the two-atom experiment
and the overall measurement time necessary to violate Bell's inequality
with sufficient statistical significance. In the current experiment,
the sequence for generation of atom-photon entanglement consists of
the preparation of the initial state by optical pumping ($\sim5\,\mathrm{\mu s}$)
and excitation. Currently after every $20$ preparation-excitation
cycles the atom has to be cooled for $200\,\mathrm{\mu s}$, which
gives additional $10\,\mathrm{\mu s}$ per cycle. For the remote entanglement
the emitted photon will be sent over an optical fiber of about $200\,\mathrm{m}$
length to the place where entanglement swapping is performed. Therefore
a waiting time of $\frac{2\cdot200\,\mathrm{m}}{\frac{2}{3}c}=2\,\mathrm{\mu s}$
is necessary to send the photon and to receive a signal about the
success or failure of the entanglement swapping procedure. This gives
altogether $17\,\mathrm{\mu s}$ per cycle and a repetition rate of
$58.8\,\mathrm{kHz}$. Assuming a mean occupation number of each trap
of $0.5$ we get the duty cycle of the two-trap system of at least
$(0.5)^{2}=0.25$. This results in an effective repetition rate of
$0.25\cdot58.8\,\mathrm{kHz}=14.7\,\mathrm{kHz}$. Together with the
success probability (\ref{eq:efficiency-EntSwap}) of the entanglement
swapping process of $2.34\cdot10^{-7}$ we expect $1$ atom-atom event
in approximately $5$ minutes. Depending on the detection method it
is necessary to evaluate between $2600$ and $3470$ atom-atom events
in order to violate Bell's inequality by $3$ standard deviations.
This requires a continuous measurement time between $9$ and $12$
days. By detection of a second Bell state during the BSM\cite{Mattle96}
this measurement time could be reduced by a factor of two.

\section{Summary}

We have shown the feasibility of a loophole-free test of Bell's inequality
with entangled pairs of neutral atoms. By simultaneously exciting
two single $^{87}Rb$ atoms in remote traps and detecting interference
of the emitted photons it should be possible to entangle the atoms
with a high fidelity. The two available methods of atomic state detection
allow to violate Bell's inequality by achieving an $S\sim2.1>2$ and
to evaluate the complete ensemble of entangled atom pairs (i.e. without
the need for a fair sampling assumption). Additionally, strict space-like
separation of measurement events is obtainable by using a distance
between the atomic traps of $300\,\mathrm{m}$. Although very challenging,
this approach is a promising candidate for a conclusive test of quantum
mechanics against theories with local hidden variables.

This work was supported by the Deutsche Forschungsgemeinschaft, the
Munich-Centre for Advanced Photonics MAP, the European Commission
through the EU Project QAP (IST-3-015848), the Elite Network of Bavaria
through the excellence program QCCC, and the Polish-German Exchange
Program DAAD-MNISW.


\begin{thebibliography}{19}
\bibitem{footnote1}It can be shown that the spatial mode filtering
given by the single-mode optical fiber leads to observation of a maximally
entangled state for an arbitrarily large numerical aperture of the
photon collecting optics.
\end{thebibliography}
\end{document}